\documentclass[a4paper,10pt,twocolumn]{article}
 
\usepackage[english]{babel}
\usepackage[utf8]{inputenc}
\usepackage[T1]{fontenc}
\usepackage{url}

\usepackage{xspace}
\usepackage[normalem]{ulem}
\usepackage{ifthen}
\usepackage{xcolor}

\usepackage[top=1.5cm, left=1.5cm, right=1.5cm, bottom=1.5cm]{geometry}

\renewenvironment{abstract}{\bf\small {\em\ Abstract---}}{}

\usepackage{amsfonts,amssymb,amsmath,amsthm}
\usepackage{subfigure}
\usepackage{graphicx}
\usepackage[footnotesize]{caption}

\usepackage{amsfonts,amssymb,amsmath,amsthm}

\newtheorem{theorem}{Theorem}[section]

\newtheorem{proposition}[theorem]{Proposition}

\theoremstyle{definition}

\newcommand{\bb}{\mathbb}
\newcommand{\bs}{\boldsymbol}
\newcommand{\cl}{\mathcal}

\newcommand{\st}{%
  \ifmmode
  \mathrm{s.\!t.\xspace}%
  \else%
  \emph{s.t.}\@\xspace%
  \fi%
}

\newcommand{\ie}{\emph{i.e.}, }
\newcommand{\eg}{\emph{e.g.}, }

\newcommand{\Rbb}{\bb{R}} 
\newcommand{\Cbb}{\bb{C}} 

\newcommand{\tinv}[1]{{\textstyle\frac{1}{#1}}}

\newcommand{\im}{\mathrm{i}\mkern1mu} 

\newcommand{\iid}{%
  \ifmmode
  \mathrm{i.i.d.}%
  \else%
  i.i.d.\@\xspace%
  \fi%
}

\newcommand{\ud}{\mathrm{d}} 

\newcommand{\norm}[3][]{#1\lVert#2#1\rVert_{#3}} 
\newcommand{\scp}[3][]{#1\langle #2, #3 #1\rangle} 

\DeclareMathOperator*{\argmin}{arg\,min}

\newcommand{\intMa}[1][]{
   \ifthenelse{ \equal{#1}{} }{\bs{\cl I}}{\cl I_{#1}}
}
\newcommand{\intMOm}[2][]{
   \ifthenelse{ \equal{#2}{} }{\bs{\cl I}_\Omega}{\bs{\cl I}_\Omega #1[ #2 #1]} 
}
\newcommand{\dintMOm}[2][]{
   \ifthenelse{ \equal{#2}{} }{\tilde{\bs{\cl I}}_\Omega}{\tilde{\bs{\cl I}}_\Omega #1[ #2 #1]} 
}
\newcommand{\intCirc}[1][]{
   \ifthenelse{ \equal{#1}{} }{\bs{\cl J}}{\cl J_{#1}} 
}
\newcommand{\dintCirc}[1][]{
   \ifthenelse{ \equal{#1}{} }{\tilde{\bs{\cl J}}}{\tilde{\cl J}_{#1}} 
}
\newcommand{\ropA}{\bs{\cl A}} 
\newcommand{\ILEop}{\bs{\cl B}} 

\newcommand{\fvign}{f^{\circ}}
\newcommand{\ts}{\textstyle}

\renewcommand{\leq}{\leqslant}
\renewcommand{\geq}{\geqslant}

\newcommand{\br}[1]{\textcolor{red}{[\textbf{BR:#1}]}}
\renewcommand{\br}[1]{}

\title{Interferometric single-pixel imaging with a multicore fiber}

\author{Olivier Leblanc$^1$, Matthias Hofer$^2$, Siddharth Sivankutty$^3$, 
Hervé Rigneault$^2$ and Laurent Jacques$^1$.\\
\footnotesize{$^1$ISPGroup, ICTEAM, UCLouvain, Belgium. 
$^2$Institut Fresnel, Marseille, France. $^3$PhLAM, Lille, France.}
} \date{\empty} 

\begin{document}

\maketitle

\begin{abstract} 
  Lensless illumination single-pixel imaging with a multicore fiber (MCF) is a computational imaging technique that enables potential endoscopic observations of biological samples at cellular scale. In this work, we show that this technique is tantamount to collecting multiple symmetric rank-one projections (SROP) of a Hermitian \emph{interferometric} matrix---a matrix encoding the spectral content of the sample image. In this model, each SROP is induced by the complex \emph{sketching} vector shaping the incident light wavefront with a spatial light modulator (SLM), while the projected interferometric matrix collects up to $O(Q^2)$ image frequencies for a $Q$-core MCF. While this scheme subsumes previous sensing modalities, such as raster scanning (RS) imaging with beamformed illumination, we demonstrate that collecting the measurements of $M$ random SLM configurations---and thus acquiring $M$ SROPs---allows us to estimate an image of interest if $M$ and $Q$ scale linearly (up to log factors) with the image sparsity level, hence requiring much fewer observations than RS imaging or a complete Nyquist sampling of the $Q \times Q$ interferometric matrix. This demonstration is achieved both theoretically, with a specific restricted isometry analysis of the sensing scheme, and with extensive Monte Carlo experiments. Experimental results made on an actual MCF system finally demonstrate the effectiveness of this imaging procedure on a benchmark image.
\end{abstract}
%
\vspace*{-0.4cm}
\section{Introduction}
\label{sec:intro}

Recently, the imaging community became more and more interested in lensless solutions, providing opportunities to design imaging systems free from the constraints imposed by traditional camera architectures. Cheaper, lighter, and enabling compressive imaging with large field-of-view (FOV), Lensless Imaging (LI) is convenient for medical applications such as microscopy \cite{microscopy1, microscopy2} and \emph{in vivo} imaging \cite{invivo1, invivo2} where the extreme miniaturization of the imaging probe (diameter $\le$ 200 $\mu$m) offers a minimally invasive route to image at depths unreachable in microscopy \cite{Boominathan2016}.
Paving the way for deep biological tissues \cite{Choi22} and brain imaging with the capability to produce focal planes at various distances from the fiber tip, intensive research effort emerged for Lensless Endoscopy (LE) using multimode fibers \cite{Septier22, Lochocki22, Psaltis2016, Cizmar2012} or MultiCore Fibers (MCF) \cite{Sivankutty2016, Andresen2016, Choudhury2019}.

In the field of Computational Imaging (CI) to which LI belongs, a mathematical model is required to describe the observations as a function of the object to be imaged. Regarding the efficiency aspects, two categories of requirement must be considered when developping CI applications; (i) the model must be physically reliable but also computationnally efficient to speed up the reconstruction algorithms (ii) the acquisition method must minimize the number of observations (also called \emph{sample complexity}) needed to accurately estimate the object of interest while remaining fast. 
In single-pixel MCF-LI, \emph{Speckle Imaging} (SI) consists in randomly shaping the wavefront of the light input to the (weakly coupled) cores entering the MCF to illuminate the entire object with a randomly distributed intensity. The fraction of the light re-emitted (either at other wavelengths by fluorescence or by simple reflection) is integrated in a single-pixel sensor, playing the role of a complete projection of the speckle on the object. 
Compared to Raster Scanning (RS) the object with a translating focused spot \cite{Sivankutty2016}, SI has been shown to reduce the sample complexity \cite{Guerit2021}.

In this work, we push a leap forward towards real-time compressive LE, keeping the low sample complexity enabled by SI and introducing light propagation physics in the forward model of MCF imaging using a wavefront-shaping device.
We point out that inserting the physics yields an interferometric sensing model similar to radio-inteferometry applications \cite{Carrillo2012, wiaux2009compressed}, where the interferences of the light emmited by the cores composing the MCF give specific access to the Fourier content of the object to be imaged. 
The sampling complexity of the underlying model is analyzed both theoretically and experimentally. 

\begin{figure}[t]
  \centering
  \includegraphics[width=.8\linewidth]{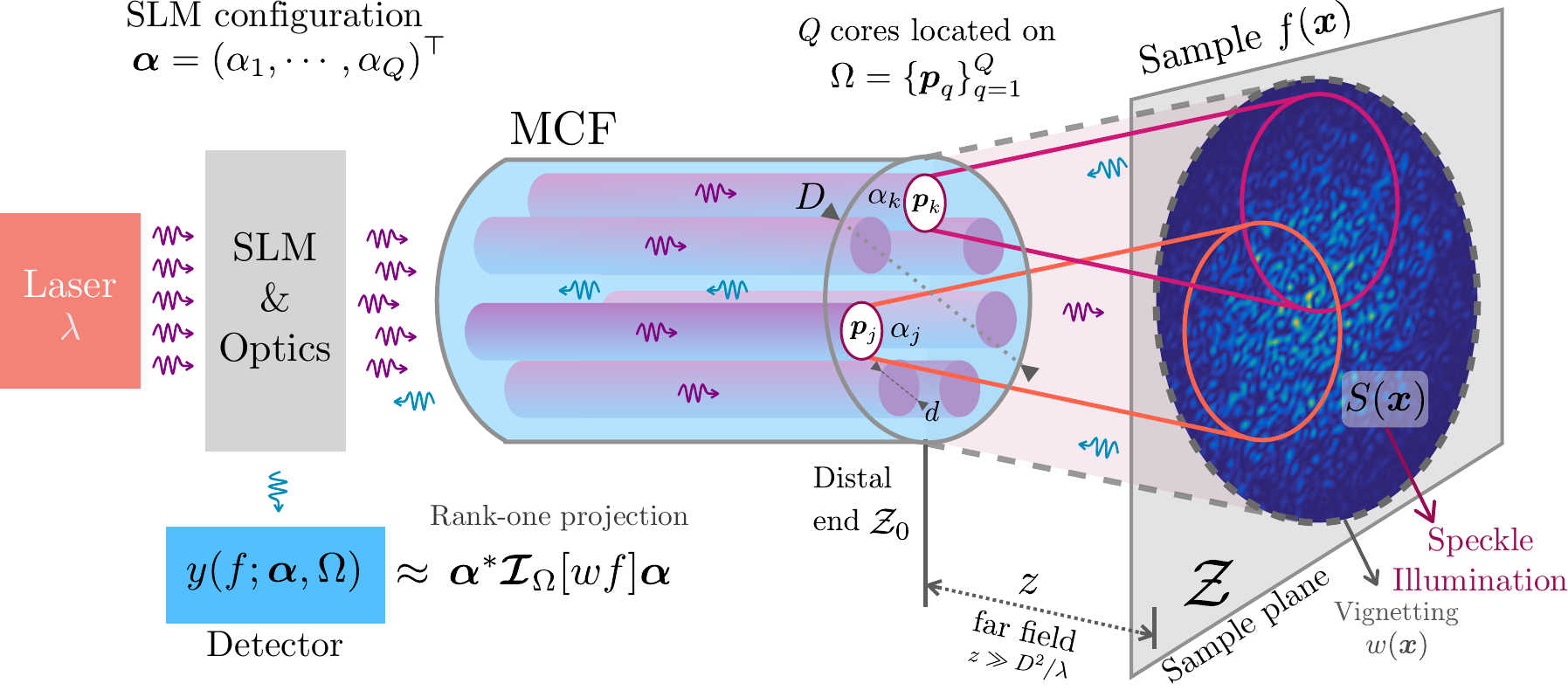}
  \\[5mm]
  \includegraphics[width=.8\linewidth]{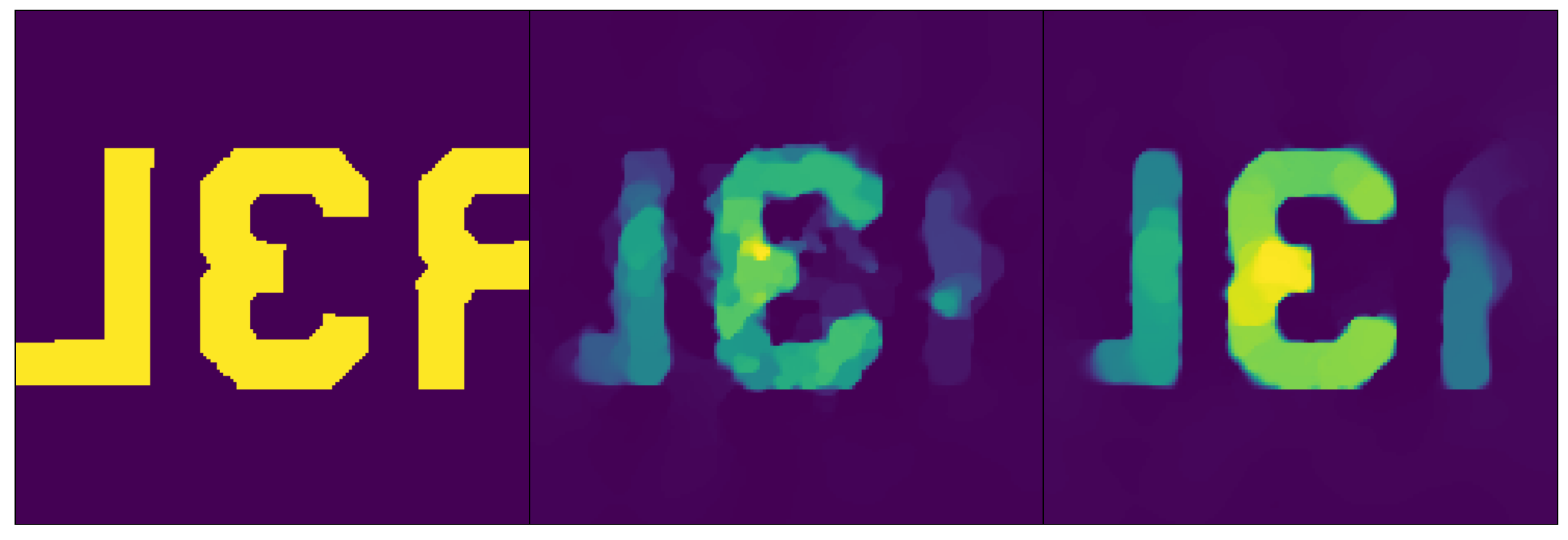}
  \caption{(top) Interferometric LI and its link with ROPs of the 
  interferometric matrix. (bottom) left: ground truth, center: reconstruction with ($Q$,$M$)=(110,49), right: reconstruction with ($Q$,$M$)=(110,$2\cdot 10^4$).}
  \label{fig:LIMCF}
\end{figure}

\section{Sensing model}
\label{sec:sensing}

Considering an MCF with diameter $D$ and $Q$ fiber cores 
(see Fig.~\ref{fig:LIMCF}-top) whose locations on the 
MCF distal end $\cl Z_0$ are in $\Omega := 
\{\bs p_q\}_{q=1}^Q \subset \Rbb^2$, and assuming a planar 
sample in the plane $\cl Z$ at a distance $z$ from $\cl Z_0$, 
by optically shaping the light wavefront with an SLM, we 
can set to $\alpha_q \in \Cbb$ the complex amplitude of 
the electromagnetic field at each fiber core $\bs p_q$. 
Writing $\bs \alpha = (\alpha_1, \ldots, \alpha_Q)^\top 
\in \Cbb^Q$ and $\bs x \in \Rbb^2$ a point on $\cl Z$, 
under the far-field approximation $z \gg D^2/\lambda$ 
(with $\lambda$ the laser wavelength) the illumination 
produced by the MCF on $\cl Z$ reads~\cite{Guerit2021}
$$
\ts S(\bs x; \bs \alpha) \approx w(\bs x)\, \big|\!\sum_{q=1}^Q \alpha_q 
e^{\frac{2\pi \im}{\lambda z} \bs p_q^\top \bs x}\big|^2,
$$
where $w$ is a smooth vignetting function---a Gaussian 
envelope with a diameter inversely proportional to the 
fiber cores diameter---determining the FOV. 
In RS mode, a focused beam can be obtained in $\cl Z$ when 
the $Q$ fiber core locations in $\Omega$ are arranged in 
a Fermat's golden spiral shape~\cite{Sivankutty2018}; 
we will restrict our analysis to this configuration. 
The LE collects a fraction $c \in (0,1)$ of the light $y$ 
globally re-emitted by the sample---as modeled by the 
fluorophore density map $f(\bs x)$---under the illumination 
$S$. For short time exposure and low intensity illumination, 
fluorescence theory provides (in a noiseless regime)
\begin{align*} 
\vspace*{-0.4cm}
&y(f; \bs \alpha, \Omega) = c \int_{\Rbb^2} 
S(\bs x; \bs \alpha) f(\bs x) \ud \bs x \\
\vspace*{-0.3cm}
&= c \sum_{j,k=1}^Q \alpha_j^* \alpha_k \int_{\Rbb^2} 
e^{\frac{2\pi \im}{\lambda z} (\bs p_k - \bs p_j)^\top \bs x} 
w(\bs x) f(\bs x) \ud \bs x \in \Rbb_+, 
\end{align*}
with the number of collected photons $y$ being Poisson 
distributed. 
Therefore, introducing the \emph{interferometry} matrix 
$\intMOm{g} \in \Cbb^{Q \times Q}$ such that, 
for a function $g: \Rbb^2 \to \Rbb$, 
$(\intMOm{g})_{jk} := \int_{\Rbb^2} 
e^{\frac{2\pi \im}{\lambda z} (\bs p_k - \bs p_j)^\top \bs x} 
g(\bs x) \ud \bs x$ with $\intMOm{g}$ Hermitian, 
assuming $c=1$ and considering the scenario where we 
collect $M$ LE observations 
$\bs y = (y_1, \ldots, y_M)^\top$, each associated 
with a specific $\bs \alpha_m$ for $1\leq m \leq M$,  
\begin{equation*}
  y(f; \bs \alpha_m, \Omega) 
  = \bs \alpha_m^* \, \intMOm{\fvign\,} \, 
  \bs \alpha_m 
  = \scp[\big]{\bs \alpha_m\bs \alpha_m^*}{ 
  \intMOm[\big]{\fvign\,}}_F,
  \label{eq:single-ROP-LE}
\end{equation*}
with $\fvign := w f$ is the image $f$ vignetted by $w$.
Under a high photon counting regime, and gathering all 
possible noise sources in an additive term $\bs n$, we can 
thus compactly write the SI sensing as
\vspace*{-0.3cm}
\begin{align} \label{eq:ROP-LE}
\begin{split} 
\bs y = \ropA \circ \intMOm[\big]{\fvign\,} + \bs n,~ 
&\text{with} (\cl A[\bs H])_m := \bs \alpha_m^* \bs{H\alpha}_m, 
\\ &\text{for}\ \bs H=\bs H^*\ \text{and}\ 
1\leq m \leq M.
\end{split}
\end{align}
We thus observe that \eqref{eq:ROP-LE} is tantamount to 
first sampling the Fourier transform of $\fvign$ over 
frequencies selected in the difference set $\cl V := 
\frac{2\pi}{\lambda z}(\Omega - \Omega) 
= \{\frac{2\pi}{\lambda z} (\bs p_j - \bs p_k)\}_{j,k=1}^Q$, 
and next performing $M$ symmetric rank-one projections 
(SROP \cite{chen2015exact,cai2015rop}) of 
$\intMOm{\fvign\,}$ as determined by $\cl A$ and 
the complex amplitude vectors $\{\bs \alpha_m\}_{m=1}^M$. 
Thus, the SI sensing corresponds to a specific 
interferometric system: assuming we collect enough SROP 
observations, we can potentially estimate the 
interferometry matrix $\intMOm{\fvign\,}$. 
The system is thus equivalent to the radio-interferometry 
principles \cite{wiaux2009compressed}---each fiber core 
playing somehow the role of a radio telescope and each 
entry of $(\intMOm{\fvign\,})_{jk}$ probing the 
frequency content of $\fvign$ on the ``\emph{visibility}'' 
$\nu_{jk} := \frac{2\pi}{\lambda z} (\bs p_j - \bs p_k)$.
%
\section{Interferometric structural models} 
\label{sec: sparsity_on_intM}
If one aims to image a vignetted sample that is $K$-sparse 
in the spatial domain, \ie it is composed of 
a few spikes as $\fvign (\bs x) = 
\sum_{i=1}^K \rho_i \delta (\bs x - \bs{x}_i)$ for 
$K \ll Q$, the entry "$jk$" of the interferometric 
matrix reads $\intMOm[\big]{\fvign}_{jk} = \sum_{i=1}^K \rho_i 
e^{\im 2\pi (\bs{p}_k - \bs{p}_j)^\top \bs{x}_i}.$
We can therefore write the interferometric matrix as
\begin{equation*} 
    \intMOm[\big]{\fvign} = \sum_i \rho_i 
    \bs u(\bs{x}_i) \bs u^*(\bs{x}_i), ~~~~ \bs{u}(\bs x)_j := 
    e^{-\im 2\pi \bs{p}_j^\top \bs x},
\end{equation*}
which shows that it is low-rank with rank $K$. 
More generally, for a sample $\fvign$ that can be assumed
sparsely represented in a collection of functions 
$\{ \psi_k\}_{k=1}^d$ (\eg a wavelet basis for bandlimited function 
supported inside $\Omega$), \ie $\fvign (\bs x) = \sum_{k=1}^d 
\rho_k \psi_k (\bs x)$ with $\norm{\bs \rho}{0}=K \ll d$, then the interferometric matrix belongs to a subspace of dimension $K$ and writes as 
$\intMOm{\fvign} = \sum_{k | \rho_k \neq 0}^K \rho_k \intMOm{\psi_k}$.

\section{Image reconstruction}
\label{sec:reconstruction}
Assuming the sample $\fvign \in \Omega$ is band-limited,
we are interested in accurately estimating a discretisation $\bs f 
\in \bb R^N$ of $\fvign$
We consider a discretisation of (\ref{eq:ROP-LE}) that reads 
\begin{equation*}
  \bs y = \ropA \circ \dintMOm{\bs f} + \bs n, \text{with } \dintMOm{\bs f} = 
  \bs R_{\tilde{\cl V}} \bs F \bs f,
\end{equation*} 
where $\bs F$ is the Fourier matrix and 
$\bs R_{\tilde{\cl V}}: \Cbb^N \mapsto \Cbb^{Q\times Q}$ is the restriction 
to the set $\tilde{\cl V}$ obtained as a Cartesian gridding of the (off-grid) 
difference set $\cl V$ (reached by nearest neighbors). 
From the factorization of this model, we first conclude 
that the set $\tilde{\cl V}$ should ideally be composed 
of as many distinct frequencies as possible (except for the 
zero frequency that has multiplicity $Q$) to improve our 
knowledge of $\bs f$. 
Interestingly, we can show numerically that Fermat's gold 
spiral arrangement ensures the unicity of the visibilities 
$\tilde\nu_{jk}$ when $j \neq k$, \ie $\tilde{\cl V}$ is composed of 
$Q(Q-1) +1$ distinct frequencies. 
In a noiseless scenario, we have shown that there 
exists a combination of $M_0 = O(Q^2)$ 
deterministic ROP observations that exactly recovers 
$\dintMOm{\fvign\,}$.   
Therefore, in a compressive setting, we could first 
leverage the low-complexity structure of 
$\dintMOm{\fvign\,}$---as induced from that of 
$\bs f$---to recover $\dintMOm{\bs f}$ from 
$M < M_0$ random complex ROPs, and then infer $\bs f$ from its $Q(Q-1) +1$
frequencies encoded in $\dintMOm{\bs f}$, this second step 
being similar to the inverse problem posed in 
radio-interferometry~\cite{wiaux2009compressed, Carrillo2012}. 
In a simpler case where the cores are placed at all integer positions, 
the interferometric matrix is shown to be circulant and low-rank,
\ie $\dintMOm{\bs f} = \cl T \circ \bs F \bs f$ where $\cl T : \Cbb^N \mapsto 
\Cbb^{N\times N}$ is the operator that turns a vector into a circulant matrix. 
In this situation, the sensing operator 
$\cl B$ respects a 
specific RIP-$\ell_2/\ell_1$ property over the set of 
sparse images---thus 
extending former approaches restricted to real sparse and 
low-rank matrices~\cite{chen2015exact}, and the RIP of 
random partial Fourier sensing characterized by 
$\tilde{\cl V}$~\cite{foucart2017mathematical}.
Proposition~\ref{prop:L2L1} shows that proving this RIP-$\ell_2/\ell_1$
for $\cl B$ implies we can reliably estimate it in a single 
basis pursuit denoising program (BPDN) with an $\ell_1$ 
fidelity term. 
This happens with high probability if the vectors 
$\{\bs \alpha_m\}_{m=1}^M$ are random and sub-Gaussian, 
and both $M$ and $Q^2$ are large compared to the sparsity 
level of $\bs f$. 

\begin{proposition}[$\ell_2/\ell_1$ instance optimality of BPDN$_{\ell_1}$]
  \label{prop:L2L1}
  Let $\ILEop := \ropA \circ \cl T \circ \bs F$ be an operator 
  that respects the $\text{RIP}_{\ell_2/\ell_1}
  (k, \alpha_k, \beta_k)$ for $k \in \{K, K+K' \}$ with $K'>2K$, and $
  \tinv{\sqrt 2} m_{K+K'} - M_{K'} \frac{\sqrt K}{\sqrt{K'}} \geq \gamma >0$, for some $M_{K'} >0$. Then, $\forall \bs f \in \Rbb^N$ 
  , the estimate 
  \begin{equation*}
      \hat{\bs f} \in \argmin_{\bs u}~ \norm{\bs u}{1} ~~\st 
      \norm[\big]{\underbrace{\ILEop(\bs f)+ \bs n}_{\bs y} - 
      \ILEop(\bs u) }{1} \leq \epsilon 
      \vspace*{-0.4cm}  
  \end{equation*}
  satisfies
  \vspace*{-0.2cm}
  \begin{equation*}
      \norm{ \bs f-\hat{\bs f}}{2} \leq C \frac{\norm{ 
        \bs f-\bs f_K}{1}}{\sqrt K} + D \frac{\epsilon}{m}
  \end{equation*}
\end{proposition}

The current theoretical derivations are accompanied by 
numerical (not shown in this abstract) and experimental reconstruction results (see lead-in in Fig.~\ref{fig:LIMCF}-bottom) that suggest Proposition~\ref{prop:L2L1} may also hold when relaxing the cited assumptions. Additionnally, this may extend to other optimisation problems like LASSO \cite{pareto} or lagrangian formulations with various regularization terms  ($\ell_1$ in the identity or orthonormal basis, total variation).
\bibliographystyle{unsrt}
\bibliography{biblio}

\begin{thebibliography}{10}

\bibitem{microscopy1}
Aydogan Ozcan and Utkan Demirci.
\newblock Ultra wide-field lens-free monitoring of cells on-chip.
\newblock {\em Lab Chip}, 8:98--106, 2008.

\bibitem{microscopy2}
Aydogan Ozcan and Euan McLeod.
\newblock Lensless imaging and sensing.
\newblock {\em Annual Review of Biomedical Engineering}, 18(1):77--102, 2016.
\newblock PMID: 27420569.

\bibitem{invivo1}
Grace Kuo, Fanglin~Linda Liu, Irene Grossrubatscher, Ren Ng, and Laura Waller.
\newblock On-chip fluorescence microscopy with a random microlens diffuser.
\newblock {\em Opt. Express}, 28(6):8384--8399, Mar 2020.

\bibitem{invivo2}
Jesse~K. Adams, Vivek Boominathan, Sibo Gao, Alex~V. Rodriguez, Dong Yan, Caleb
  Kemere, Ashok Veeraraghavan, and Jacob~T. Robinson.
\newblock In vivo fluorescence imaging with a flat, lensless microscope.
\newblock {\em bioRxiv}, 2020.

\bibitem{Boominathan2016}
Vivek Boominathan, Jesse~K. Adams, M.~Salman Asif, Benjamin~W. Avants, Jacob~T.
  Robinson, Richard~G. Baraniuk, Aswin~C. Sankaranarayanan, and Ashok
  Veeraraghavan.
\newblock {Lensless Imaging: A computational renaissance}.
\newblock {\em IEEE Signal Processing Magazine}, 33(5):23--35, 2016.

\bibitem{Choi22}
W.~Choi, M.~Kang, and J.H. et~al. Hong.
\newblock Flexible-type ultrathin holographic endoscope for microscopic imaging
  of unstained biological tissues.
\newblock {\em Nature Communications}, 4469(13), 2022.

\bibitem{Septier22}
D.~Septier, V.~Mytskaniuk, R.~Habert, D.~Labat, K.~Baudelle, A.~Cassez,
  G.~Br\'{e}valle-Wasilewski, M.~Conforti, G.~Bouwmans, H.~Rigneault, and
  A.~Kudlinski.
\newblock Label-free highly multimodal nonlinear endoscope.
\newblock {\em Opt. Express}, 30(14):25020--25033, Jul 2022.

\bibitem{Lochocki22}
Benjamin Lochocki, Max V.~Verweg, Jeroen J.~M.~Hoozemans, Johannes F.~de Boer,
  and Lyubov V.~Amitonova.
\newblock Epi-fluorescence imaging of the human brain through a multimode
  fiber.
\newblock {\em APL Photonics}, (7), 2022.

\bibitem{Psaltis2016}
Demetri Psaltis and Christophe Moser.
\newblock {Imaging with Multimode Fibers}.
\newblock {\em Optics and Photonics News}, 27(1):24----31, 2016.

\bibitem{Cizmar2012}
Tomas Cizmar and Kishan Dholakia.
\newblock {Exploiting multimode waveguides for pure fibre-based imaging}.
\newblock {\em Nature Communications}, 3(May), 2012.

\bibitem{Sivankutty2016}
Siddharth Sivankutty, Viktor Tsvirkun, G{\'{e}}raud Bouwmans, Dani Kogan, Dan
  Oron, Esben~Ravn Andresen, and Herv{\'{e}} Rigneault.
\newblock {Extended field-of-view in a lensless endoscope using an aperiodic
  multicore fiber}.
\newblock {\em Optics Letters}, 41(15):3531, 2016.

\bibitem{Andresen2016}
Esben~Ravn Andresen, Siddharth Sivankutty, Viktor Tsvirkun, G{\'{e}}raud
  Bouwmans, and Herv{\'{e}} Rigneault.
\newblock {Ultrathin endoscopes based on multicore fibers and adaptive optics:
  status and perspectives}.
\newblock {\em Journal of Biomedical Optics}, 21(12):121506, 2016.

\bibitem{Choudhury2019}
Debaditya Choudhury, Duncan~K. McNicholl, Audrey Repetti, Itandehui
  Gris-S{\'{a}}nchez, Tim~A. Birks, Yves Wiaux, and Robert~R. Thomson.
\newblock {Compressive optical imaging with a photonic lantern}.
\newblock 2019.

\bibitem{Guerit2021}
St{\'e}phanie Gu{\'e}rit, Siddharth Sivankutty, John Lee, Herv{\'e} Rigneault,
  and Laurent Jacques.
\newblock Compressive imaging through optical fiber with partial speckle
  scanning.
\newblock {\em SIAM Journal on Imaging Sciences}, 15(2):387--423, 2022.

\bibitem{Carrillo2012}
R.~E. Carrillo, J.~D. McEwen, and Y.~Wiaux.
\newblock {Sparsity Averaging Reweighted Analysis (SARA): A novel algorithm for
  radio-interferometric imaging}.
\newblock {\em Monthly Notices of the Royal Astronomical Society},
  426(2):1223--1234, 2012.

\bibitem{wiaux2009compressed}
Yves Wiaux, Laurent Jacques, Gilles Puy, Anna~MM Scaife, and Pierre
  Vandergheynst.
\newblock Compressed sensing imaging techniques for radio interferometry.
\newblock {\em Monthly Notices of the Royal Astronomical Society},
  395(3):1733--1742, 2009.

\bibitem{Sivankutty2018}
Siddharth Sivankutty, Viktor Tsvirkun, Olivier Vanvincq, G{\'{e}}raud Bouwmans,
  Esben~Ravn Andresen, and Herv{\'{e}} Rigneault.
\newblock {Nonlinear imaging through a Fermat's golden spiral multicore fiber}.
\newblock {\em Optics Letters}, 43(15):3638, 2018.

\bibitem{chen2015exact}
Yuxin Chen, Yuejie Chi, and Andrea~J Goldsmith.
\newblock Exact and stable covariance estimation from quadratic sampling via
  convex programming.
\newblock {\em IEEE Transactions on Information Theory}, 61(7):4034--4059,
  2015.

\bibitem{cai2015rop}
T~Tony Cai, Anru Zhang, et~al.
\newblock Rop: Matrix recovery via rank-one projections.
\newblock {\em Annals of Statistics}, 43(1):102--138, 2015.

\bibitem{foucart2017mathematical}
Simon Foucart and Holger Rauhut.
\newblock A mathematical introduction to compressive sensing.
\newblock {\em Bull. Am. Math}, 54(2017):151--165, 2017.

\bibitem{pareto}
D.~E.N. {Van Ewout Berg} and Michael~P. Friedlander.
\newblock {Probing the pareto frontier for basis pursuit solutions}.
\newblock {\em SIAM Journal on Scientific Computing}, 31(2):890--912, 2008.

\end{thebibliography}
\end{document}